\documentclass[aps,prd,twocolumn,superscriptaddress,nofootinbib,floatfix]{revtex4-2}
\usepackage{amsmath,amssymb,graphicx,booktabs}
\usepackage{appendix}
\usepackage{orcidlink}
 \usepackage{hyperref} 
\newcommand{\Lz}{\Lambda^{(0)}}
\newcommand{\Lt}{\Lambda^{(2)}}

\begin{document}

\title{ Universal Relations Meet Causality: \\
\vskip 0.2cm
Bounding the Dynamical Love Number of Neutron Stars}

\author{Alex Kehagias\orcidlink{0000-0001-6080-6215}}
\email{kehagias@central.ntua.gr}
\affiliation{Physics Division, National Technical University of Athens, 15780 Zografou Campus, Athens, Greece}

\author{Antonio Riotto\orcidlink{0000-0001-6948-0856}}
\email{antonio.riotto@unige.ch}
\affiliation{D\'epartement de Physique Th\'eorique and Gravitational Wave Science Center, Universit\'e de Gen\`eve, 24 quai E.~Ansermet, CH-1211 Gen\`eve 4, Switzerland}

\begin{abstract}
\noindent
The frequency-dependent tidal response of a neutron star is constrained by causality, reality and passivity, which turn a suitably rescaled response into a holomorphic self-map of the upper complex frequency half-plane after some suitable conditions. The Schwarz--Pick theorem then bounds the leading dynamical correction to the tidal deformability from above. In this work we confront this bound with the most recent calculations of the dynamical tidal response in the literature which have also shown that the static tidal
deformability and its correction at second order in frequency  are  tightly correlated. We find that the  bound is saturated to within a few per cent, a quantitative confirmation of f-mode dominance in the relativistic regime. The combination of the static deformability with an independent determination of the f-mode frequency thus may provide an equation-of-state-independent upper limit on the dynamical tide.

\end{abstract}

\maketitle

\section{Introduction}
\noindent
A compact object such as a neutron star, when immersed in an external gravitational field which varies across its size, gets deformed and develops induced multipole moments. The coefficients relating the induced moments to the applied tidal field are the tidal Love numbers. They leave an imprint on the gravitational wave (GW) signal emitted by a coalescing binary~\cite{Flanagan:2007ix,Hinderer:2007mb} and have become one of the cleanest probes of the internal structure of the inspiralling bodies~\cite{Flanagan:2007ix,Rodriguez:2026iot,Chakraborty:2026qru}. A static tide induces a constant moment. In a realistic inspiral, however, the tide varies in time and the induced moment acquires a frequency dependence, whose low-frequency expansion defines the dynamical Love numbers.

Let us denote by $M$, $R$ and $C=M/R$ the mass, radius and compactness of the star, respectively. For a perfect (non-dissipative) fluid, the quadrupolar tidal deformability at angular frequency $\omega$ reads~\cite{Saes:2025jvr}
\begin{equation}\label{eq:Lambda}
\Lambda(\omega)=\frac{2k_2(\omega)}{3C^5}=\Lz+\Lt(M\omega)^2+\dots ,
\end{equation}
where $k_2(\omega)$ is the frequency-dependent Love number and $\omega$ is the angular frequency of the tidal field, which coincides with the GW angular frequency for the dominant $\ell=m=2$ harmonic. The ratio of the two coefficients defines an effective frequency
\begin{equation}\label{eq:wstar}
(M\omega_*)^2\equiv\frac{\Lz}{\Lt} ,
\end{equation}
i.e.\ the frequency of the single mode which reproduces both coefficients. The corresponding one-mode (f-mode) response
\begin{equation}
\Lambda_{\rm OM}(\omega)=\Lz/(1-\omega^2/\omega_*^2)
\end{equation}
is widely used in waveform modelling~\cite{Steinhoff:2016rfi,Hinderer:2016eia} and is well supported by perturbative calculations~\cite{Apostolidis:2026qsg}.

Two questions naturally arise. The first, of theoretical nature, is whether the coefficients $\Lz$ and $\Lt$ are arbitrary or are restricted by general principles. The second, of more phenomenological nature, is how sensitive they are to the details of the microphysics, in particular to the equation of state (EoS) of the star.

The first question has been recently answered in Ref.~\cite{Kehagias:2026opd}, where it has been shown that $\Lt$ is indeed constrained by general principles. The tidal response is a retarded Green's function. It is analytic in the upper half of the complex frequency plane, real on the imaginary axis and, since the star absorbs rather than supplies energy, its absorptive part has a definite sign. These properties turn a suitably rescaled response into a holomorphic map of the upper half-plane into itself. The Schwarz--Pick theorem, the same geometric statement underlying the chaos bound for thermal correlators~\cite{Maldacena:2015waa,Giataganas:2026ctn}, then states that such a map cannot stretch hyperbolic distances. Along the imaginary frequency axis this becomes a bound on the logarithmic slope of the response and, for a star whose spectral weight starts at the fundamental mode, a sharp upper bound on $\Lt$~\cite{Kehagias:2026opd}.

As for the second question, a handle on the unknown EoS is provided by quasi-universal relations, which link neutron star observables in a way that is largely insensitive to the microphysics. By removing most of the EoS dependence from macroscopic quantities, they have become a standard tool of neutron star phenomenology, the best-known examples being the I-Love-Q and the f-mode--Love relations~\cite{Yagi:2013awa,Yagi:2016bkt}. The same idea has been recently applied to dynamical tides~\cite{Saes:2025jvr} (see also Ref.~\cite{Apostolidis:2026qsg}), where for compactness values down to $C\sim 0.12$, the static tidal deformability and its leading dynamical correction, at second order in frequency, turn out to be tightly correlated.

A question then naturally follows: is such a correlation consistent with the bound dictated by general principles~\cite{Kehagias:2026opd}? Answering it is timely, as it may help to identify the regions of parameter space which are allowed (or excluded) by causality and passivity, and to explore a broader class of EoSs. This is the goal of the present paper. In Section~\ref{sec:bound} we briefly summarise the derivation of the bound, while in Section~\ref{sec:shy} we confront it with the two most recent fully relativistic computations of the dynamical tidal response. Section~\ref{sec:concl} contains our conclusions, and Appendix~\ref{AppA} a discussion of the refined bound which applies in the presence of the renormalization group running.

\section{The Schwarz--Pick bound}\label{sec:bound}
\noindent
Let us briefly summarise how the bound on the dynamical Love number of neutron stars is derived~\cite{Kehagias:2026opd}.

Three physical properties fix the analytic structure of $\Lambda(\omega)$:
\begin{itemize}
    \item
causality makes it analytic in the upper half-plane, ${\rm Im}\,\omega>0$; 
\item reality makes it real on the positive imaginary axis, $\omega=i\Omega$ with $\Omega>0$; 
\item passivity, i.e.\ the statement that over a cycle the star can only absorb energy from the tide, requires $\omega\,{\rm Im}\,\Lambda(\omega)\ge0$ for real $\omega$. 
\end{itemize}
Supplemented by a single subtraction at large frequency, these properties lead to a dispersion relation with non-negative spectral measure $d\rho$
\begin{equation}\label{eq:dispH}
\Lambda(\omega)=\Lambda^{(\infty)}+\int\frac{\omega'^2}{\omega'^2-\omega^2}\,d\rho(\omega'),\qquad d\rho\ge0,
\end{equation}
where $\Lambda^{(\infty)}\ge0$ is the instantaneous part of the response. We will set it to zero from now on based on the physical principle that  for external perturbations of sufficiently high frequency the star does not have the time to build up a coherent response.

Expanding at small frequency, the two coefficients of Eq.~\eqref{eq:Lambda} are identified with two moments of the same positive measure
\begin{equation}\label{eq:moments}
\Lz(0)=%\Lambda^{(\infty)}+
\int d\rho\geq 0,\qquad \Lt(0)=\frac{1}{M^2}\int\frac{d\rho}{\omega'^2}.
\end{equation}
The first relation tells us that our considerations apply away from the strong-gravity regime, as it is well known that for black holes the static Love number vanishes~\cite{Kehagias:2024rtz,Combaluzier--Szteinsznaider:2025eoc,Gounis:2024hcm}. Furthermore, since $\Lz$ scales like $C^{-5}$ and $\Lt$ like $C^{-8}$, the low-frequency expansion is better behaved in the small-compactness, Newtonian, regime, while relativistic effects become more and more relevant as $C$ grows.

The response itself does not map the upper half-plane $\mathbb{H}$ into itself, since its imaginary part is odd in $\omega$. The rescaled function $F(\omega)\equiv\omega\Lambda(\omega)$, however, does. Writing $\omega=x+iy$ with $y>0$, Eq.~\eqref{eq:dispH} gives
\begin{equation}\label{eq:ImF}
{\rm Im}\,F=\Lambda^{(\infty)}\,y+\int\frac{\omega'^2\,y\,\big(\omega'^2+|\omega|^2\big)}{\big|\omega'^2-\omega^2\big|^2}\,d\rho(\omega')\ge0 ,
\end{equation}
so that $F$ is a Herglotz function. On the imaginary axis it is purely imaginary, $F(i\Omega)=i\Phi(\Omega)$ with $\Phi(\Omega)\equiv\Omega\Lambda(i\Omega)>0$, and therefore it maps the positive imaginary axis into itself.

Let us now equip $\mathbb{H}$ with the hyperbolic metric $ds=|dw|/{\rm Im}\,w$. The Schwarz--Pick theorem states that a holomorphic self-map of $\mathbb{H}$ cannot increase hyperbolic distances, and preserves them only if it is a M\"obius automorphism. The positive imaginary axis is a geodesic along which $ds=d\Omega/\Omega$, so that the hyperbolic distance between $i\Omega_1$ and $i\Omega_2$ is $|\ln(\Omega_1/\Omega_2)|$. Since $F$ maps this geodesic into itself, the theorem implies $|\ln[\Phi(\Omega_1)/\Phi(\Omega_2)]|\le|\ln(\Omega_1/\Omega_2)|$, whose local version is the master inequality~\cite{Kehagias:2026opd}
\begin{equation}\label{eq:SP}
-1\le\frac{d\ln\big[\Omega\,\Lambda(i\Omega)\big]}{d\ln\Omega}\le1 .
\end{equation}
Equivalently, $\Lambda(i\Omega)$ never increases with $\Omega$, while $\Omega^2\Lambda(i\Omega)$ never decreases. The inequality constrains the full response at every Euclidean frequency. Notice that a single mode saturates it only asymptotically, for $\Omega\ll\omega_*$ or $\Omega\gg\omega_*$.

\subsection{Content of the low-frequency expansion}
\noindent
Plugging Eq.~\eqref{eq:Lambda} into Eq.~\eqref{eq:SP}, with $\Lambda(i\Omega)=\Lz-\Lt(M\Omega)^2+{\cal O}(\Omega^4)$, one finds
\begin{equation}\label{eq:slopeTaylor}
\frac{d\ln\big[\Omega\,\Lambda(i\Omega)\big]}{d\ln\Omega}=\frac{1-3u}{1-u},\qquad u\equiv\frac{\Lt}{\Lz}(M\Omega)^2=\frac{\Omega^2}{\omega_*^2}.
\end{equation}
The upper side of Eq.~\eqref{eq:SP} requires $\Lt\ge0$, a frequency-independent constraint stating that dynamical effects can only enhance the static deformability at low frequency. The lower side requires $u\le1/2$, i.e.
\begin{equation}
   \Lt\le\Lz/[2(M\Omega)^2]. 
   \label{f}
\end{equation}
This condition is necessarily violated for sufficiently large $\Omega$.  This is not surprising as we are doing an expansion in small frequencies and higher-order terms should be  included. Furthermore, in 
the effective field theory (EFT) approach to calculate the Love numbers, the procedure is valid only below a certain energy scale and the violation of Eq. (\ref{f}) 
would only signal the breakdown of the EFT. 

For a cold, non-rotating, barotropic star the lowest quadrupolar oscillation with non-vanishing tidal overlap is the f-mode, so that $d\rho=0$ for $\omega'<\omega_f$. Then $1/\omega'^2\le1/\omega_f^2$ inside Eq.~\eqref{eq:moments}, and one obtains~\cite{Kehagias:2026opd}
\begin{equation}\label{eq:mom}
\Lt\le\frac{\Lz}{(M\omega_f)^2}
\quad\Longleftrightarrow\quad \omega_*\ge\omega_f .
\end{equation}
This result has a simple interpretation. The effective frequency $\omega_*$ is a weighted average over the spectrum and cannot lie below its lowest frequency. The bound is saturated when all the weight sits at $\omega_f$, i.e.\ by the one-mode response with $\omega_*=\omega_f$, while moving weight above the f-mode lowers $\Lt$ below its maximum. If the full response is known at one Euclidean frequency, the same argument provides the two-sided bracket~\cite{Kehagias:2026opd}
\begin{equation}\label{eq:bracket}
\frac{\Lz-\Lambda(i\Omega)}{(M\Omega)^2}\le\Lt\le\Big(1+\frac{\Omega^2}{\omega_f^2}\Big)\frac{\Lz-\Lambda(i\Omega)}{(M\Omega)^2},
\end{equation}
whose width is at most a factor of two for $\Omega\le\omega_f$. Given only $(\Lz,\Lt)$, Eq.~\eqref{eq:mom} exhausts the content of the Schwarz--Pick bound, namely whenever it is satisfied, there exists a full response obeying Eq.~\eqref{eq:SP} and reproducing both coefficients, the one-mode response being an example.

\section{Quasi-universal relations}\label{sec:shy}
\noindent
After this theoretical digression, we now turn to the phenomenology and confront the bound~\eqref{eq:mom} with the most recent computations of the dynamical Love number of neutron stars.

\subsection{First comparison}
\noindent
Ref.~\cite{Saes:2025jvr}  computes $\Lz$ and $\Lt$ for 59 EoSs, 35 piecewise polytropes and 24 tabulated models with non-trivial features in the speed of sound, along sequences ranging from $C=0.12$ to the maximum mass. The results are fitted to
\begin{equation}\label{eq:fit}
\ln\Lt=\sum_{i=0}^{3}a_i\,(\ln\Lz)^i ,\qquad 2.5\lesssim\Lz\lesssim3500 ,
\end{equation}
where the coefficients $a_i$ are determined numerically as follows. For each EoS, the Tolman--Oppenheimer--Volkoff equations are integrated for a sequence of central pressures. The perturbation equations for the metric function $H$ and for the fluid displacements $W$ and $V$ are expanded as $X(r,\omega)=X^{(0)}(r)+\omega^2X^{(2)}(r)$, so that $H^{(0)}$ obeys the static tidal equation, $W^{(0)}$ and $V^{(0)}$ are sourced by it, and $H^{(2)}$ obeys an inhomogeneous equation sourced by all three. The interior solution is started from series expansions regular at the centre, subject to the vanishing of the Lagrangian pressure perturbation at the surface, and integrated from both ends to an intermediate matching radius. Outside the star the solution is written as $H\propto k_2(\omega)H_{\rm multipole}+H_{\rm tide}$, with both functions known in closed form to ${\cal O}(M^2\omega^2)$, and the continuity of $H$ and $H'$ at $r=R$ gives $k_2^{(0)}$ and $k_2^{(2)}$, hence $\Lz=2k_2^{(0)}/(3C^5)$ and $\Lt=2k_2^{(2)}/(3C^5M^2)$. The points from all EoSs are then pooled and fitted with Eq.~\eqref{eq:fit} where individual stars scatter around the fit by less than $3\%$ in $90\%$ of the cases, and by at most $5\%$.

Two sets of coefficients are provided, which differ only in how $\Lt$ is read off from the exterior field. Outside the star the static field is a combination of a solution growing at large $r$, representing the applied tide, and of a decaying one, representing the response. At ${\cal O}(\omega^2)$ each of them is corrected by a particular solution of the inhomogeneous exterior equation, which is defined only up to the addition of homogeneous solutions whose tails again grow or decay like the tide or the response. How much of the ${\cal O}(\omega^2)$ field is designated as tide rather than response remains a matter of convention, as reassigning the growing part to the tidal moment leaves the induced moment unchanged while shifting $k_2^{(2)}$ by a term proportional to $k_2^{(0)}$~\cite{Apostolidis:2026qsg}, i.e.
\begin{equation}\label{eq:shift}
\Lt\to\Lt+c\,\Lz+c',
\end{equation}
with EoS-independent numbers $c$ and $c'$. In the Pitre--Poisson (PP) convention~\cite{Pitre:2023xsr}, adopted in Appendix~D of Ref.~\cite{Saes:2025jvr}, the particular solutions take a specific closed form and the second-order tidal constant is set to zero, so that the tidal moment is frequency independent. The main analysis of Ref.~\cite{Saes:2025jvr} adopts instead the exterior functions of Ref.~\cite{HegadeKR:2024agt}. The corresponding coefficients are
\begin{align}
a^{\rm main}&=(3.446,\,0.9876,\,5.811\times10^{-2},\,-2.174\times10^{-3}),\\
a^{\rm PP}&=(2.140,\,1.267,\,4.651\times10^{-2},\,-2.652\times10^{-3}).
\end{align}
The two fits differ by $\Lt_{\rm main}-\Lt_{\rm PP}\simeq(21\pm1.5)\Lz$, uniformly over three decades in $\Lz$, which is precisely of the form~\eqref{eq:shift}. Neither choice is more physical than the other. GW observables depend only on the full field, provided the tidal field entering the orbital dynamics is defined consistently.

Let us also notice that no renormalization scale appears in Ref.~\cite{Saes:2025jvr}. In general relativity the ${\cal O}(\omega^2)$ response is accompanied by a logarithm, generated by the scattering of the tidal field off the curved exterior geometry, which in an EFT treatment appears as $\ln(\mu r)$ and makes $\Lt$ run with $\mu$ (see Refs.~\cite{Rodriguez:2026iot,Chakraborty:2026qru} for recent reviews). Working directly with the full exterior solution, Ref.~\cite{Saes:2025jvr} never introduces a regulator. The logarithms enter the exterior functions as $\ln(1-z)=\ln(2M/r)$, with $z=1-2M/r$, so that the reference scale is set by the stellar mass itself. The freedom associated to $\mu$ is instead hidden in the choice of the particular solution, since a change of $\ln\mu$ shifts $\Lt$ exactly as in Eq.~\eqref{eq:shift}. Each normalization therefore corresponds to a fixed effective scale.

To evaluate the bound~\eqref{eq:mom} we take the f-mode frequency from the quasi-universal f--Love relation~\cite{Chan:2014kua}, $M\omega_f=\sum_{i=0}^4\bar a_i(\ln\Lz)^i$ with $\bar a=(0.1820,-6.836\times10^{-3},-4.196\times10^{-3},5.215\times10^{-4},-1.857\times10^{-5})$, which is accurate to about $1\%$.

\begin{figure}[t]
\includegraphics[width=\columnwidth]{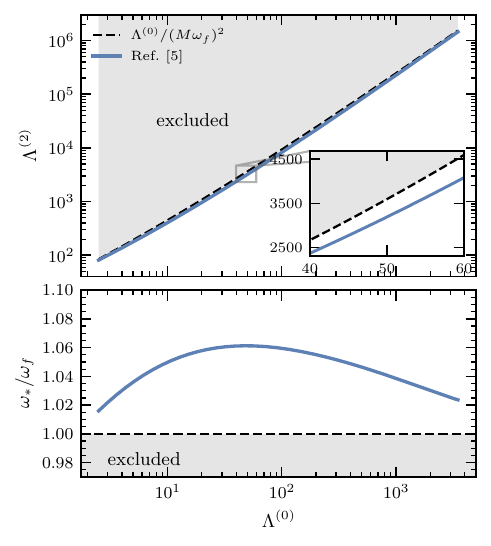}
\caption{The quasi-universal relation of Ref.~\cite{Saes:2025jvr} (solid) compared with the moment bound~\eqref{eq:mom} (dashed). Top: $\Lt$ as a function of $\Lz$, with an inset zooming on $40<\Lz<60$ in linear axes, where the gap is largest. Bottom: the ratio $\omega_*/\omega_f$. Grey regions are excluded.}
\label{fig:1}
\end{figure}

\begin{table}[b]
\caption{Ratio $\omega_*/\omega_f$ along the two quasi-universal relations of Ref.~\cite{Saes:2025jvr}.}
\label{tab:1}
\begin{ruledtabular}
\begin{tabular}{lcccccc}
$\Lz$ & 2.5 & 10 & 100 & 400 & 1000 & 3000\\
\hline
$M\omega_f$ & 0.173 & 0.150 & 0.104 & 0.079 & 0.064 & 0.050\\
main & 1.02 & 1.05 & 1.06 & 1.05 & 1.04 & 1.03\\
PP & 1.73 & 1.51 & 1.24 & 1.13 & 1.09 & 1.06\\
\end{tabular}
\end{ruledtabular}
\end{table}

As shown in Table~\ref{tab:1}, both relations lie inside the allowed region over their full range: $\Lt$ reaches $89$--$97\%$ of its maximum allowed value, i.e.\ $1.02\le\omega_*/\omega_f\le1.06$. This indicates that, despite relativistic corrections, the dynamical tidal response of neutron stars is still largely controlled by the f-mode. In the language of Eq.~\eqref{eq:moments}, almost all the spectral weight sits at the f-mode, with only a few per cent spread to higher frequencies. This is the relativistic counterpart of the Newtonian statement that the f-mode dominates the dynamical tide, and explains why the one-mode response performs well in Ref.~\cite{Saes:2025jvr}. The margin is smallest for the most compact stars, $\Lz\lesssim5$, where it becomes comparable to the scatter of individual EoSs around the fit, and a test at this level would require the individual data. The PP relation assigns a smaller $\Lt$ to the same stars and satisfies the bound with a margin which grows with compactness, reaching $\omega_*/\omega_f\simeq1.7$. In spectral terms, the same data are compatible with up to $\sim65\%$ of the weight lying well above the f-mode.
\begin{figure*}[t]
\includegraphics[width=\textwidth]{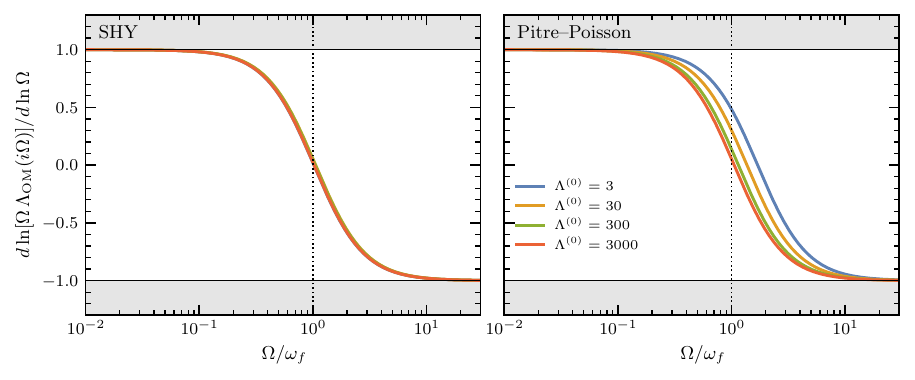}
\caption{Schwarz--Pick slope~\eqref{eq:SP} of the one-mode response built from the main-text (left) and PP (right) relations of Ref.~\cite{Saes:2025jvr}, for several values of $\Lz$. The zero crossing sits at $\Omega=\omega_*$, and the moment bound requires it to lie at or to the right of the dotted line $\Omega=\omega_f$. Grey regions are excluded.}
\label{fig:2}
\end{figure*}
We stress that the bound~\eqref{eq:SP} constrains the whole Euclidean response, and not only its first two coefficients. For the one-mode response the slope is $(1-\Omega^2/\omega_*^2)/(1+\Omega^2/\omega_*^2)$, which decreases from $+1$ to $-1$, crossing zero at $\Omega=\omega_*$, and touches the boundaries of the allowed band only asymptotically, see Fig.~\ref{fig:2}. The moment bound has a simple reading in this figure in that the zero crossing must occur at or beyond $\Omega=\omega_f$. For the main-text relation all curves cross essentially at $\omega_f$, while for the PP relation the crossing moves to higher frequencies as the compactness increases.

\subsection{Second comparison}
\noindent
Ref.~\cite{Apostolidis:2026qsg} has computed the dynamical Love number by matching relativistic stellar perturbation theory to the worldline EFT in dimensional regularization, which is needed to handle the ultraviolet divergences. The dimensionless couplings coincide with $\Lz$ and $\Lt$ of Eq.~\eqref{eq:Lambda}, but $\Lt$ runs with the renormalization scale $\mu$
\begin{align}\label{eq:run}
\Lt(\mu)&=\Lt(r_s^{-1})+\beta\ln(\mu r_s),\nonumber\\
\beta&=-\frac{64}{45}\Big(1+\frac{321}{112}\Lz\Big)<0,
\end{align}
with $r_s=2M$, so that a quasi-universal relation exists only at fixed $\mu$. Ref.~\cite{Apostolidis:2026qsg} provides it at two scales, with $X=\ln\Lz$ and $Y=\ln\Lt$,
\begin{align}\label{eq:adgfit}
Y\big|_{\mu r_s=0.01}&=3.9171+1.0707X+0.0231X^2,\nonumber\\
Y\big|_{\mu r_s=1}&=3.3840+1.1621X+0.01918X^2 ,
\end{align}
accurate to better than $10\%$ over $5\lesssim\Lz\lesssim10^5$ and consistent with Eq.~\eqref{eq:run} to about $1\%$. We use them for $5\le\Lz\le10^4$, where the f--Love relation is reliable.

\begin{figure}[t]
\includegraphics[width=\columnwidth]{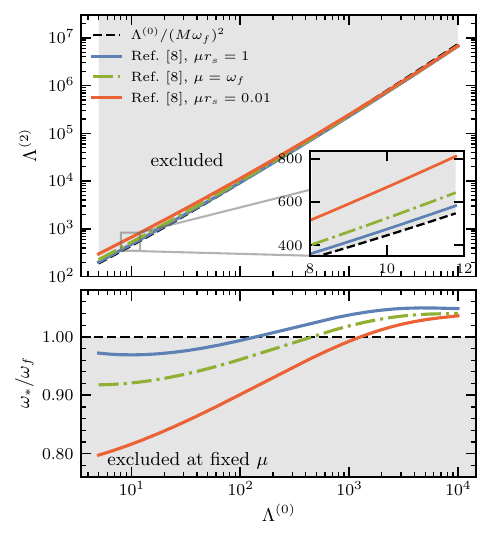}
\caption{Quasi-universal relations of Ref.~\cite{Apostolidis:2026qsg} at $\mu r_s=1$ and $\mu r_s=0.01$, together with the former evolved with Eq.~\eqref{eq:run} to $\mu=\omega_f$, compared with the moment bound~\eqref{eq:mom} (dashed). Top: $\Lt$ as a function of $\Lz$, with a zoom on $8<\Lz<12$. Bottom: $\omega_*/\omega_f$. Grey regions are excluded by the bound applied at fixed $\mu$.}
\label{fig:3}
\end{figure}
%\begin{table}[b]
%\caption{Ratio $\omega_*/\omega_f$ along the Ref. \cite{Apostolidis:2026qsg} relations of Fig.~\ref{fig:3}.}
%\label{tab:2}
%\begin{ruledtabular}
%\begin{tabular}{lcccccc}
%$\Lz$ & 5 & 30 & 100 & 300 & 1000 & $10^4$\\
%\hline
%$M\omega_f$ & 0.162 & 0.128 & 0.104 & 0.084 & 0.064 & 0.037\\
%$\mu r_s=1$ & 0.97 & 0.98 & 0.99 & 1.02 & 1.04 & 1.05\\
%$\mu=\omega_f$ & 0.92 & 0.94 & 0.96 & 0.99 & 1.02 & 1.04\\
%$\mu r_s=0.01$ & 0.80 & 0.85 & 0.90 & 0.95 & 0.99 & 1.04\\
%\end{tabular}
%\end{ruledtabular}
%\end{table}
%The comparison is shown in Fig.~\ref{fig:3} and Table~\ref{tab:2}. At $\mu r_s=1$ the relation lies at the bound: the fits give $0.97\le\omega_*/\omega_f\le1.05$, falling below unity by at most $3\%$ for $\Lz\lesssim200$, within the combined accuracy of the two fits of about $5\%$ in $\omega_*/\omega_f$. The direct computation of Ref. \cite{Apostolidis:2026dtr} for a polytrope, with $\omega_f$ the exact quasinormal-mode frequency (their Fig.~5), gives $\omega_*/\omega_f=1.015$--$1.03$ and satisfies the bound. At $\mu r_s=0.01$, on the contrary, the relation crosses into the excluded region for $\Lz\lesssim10^3$, reaching $\omega_*/\omega_f\simeq0.80$ for the most compact stars, well beyond the uncertainties and consistently with their Fig.~5, where the ratio drops to $0.95$ at $C\simeq0.18$. 
The scale $\mu$ is to be identified with the typical inverse length scale of the binary, i.e.\ it scales like the inverse of the orbital separation. In the late inspiral one expects $\mu r_s={\cal O}(10^{-1})$. One immediately notices from Fig.~\ref{fig:3} that the bound~\eqref{eq:mom} does not hold at every $\mu$, since due to  the running, $\Lt(\mu)$ grows without limit as $\mu$ decreases, while $\Lz$ and $\omega_f$ stay fixed. Such a violation is not surprising. The renormalization group running of the Love numbers becomes non-negligible for more compact objects and, as for black holes, the bound is not expected to hold when strong-gravity effects are relevant. 

By fixing the relevant physical frequency of the tidal force, one can therefore derive a cut-off on the scale $\mu$, see Fig. \ref{fig:3}. Alternatively, by fixing $\mu$, one obtains an upper bound on the frequency, to be identified with the ultraviolet cut-off of the EFT.

There is, however, a more physical way to look at the issue. The $\mu$-dependence in Eq.~\eqref{eq:run} must cancel against a logarithm of a physical scale in the full response. In the near-zone computation of Ref.~\cite{Apostolidis:2026qsg} this scale is the radial distance, through $\ln(\mu r)$, which after matching to the far zone, where the tail of the tidal field is generated, it becomes the frequency. Once a term $\omega^2\ln|\omega|$ is present at real frequencies, the only form of the full retarded response compatible with causality and reality is
\begin{equation}\label{eq:chilog}
\Lambda(\omega)=\Lz+(M\omega)^2\Big[\Lt(\mu)+\beta\ln\frac{-i\omega}{\mu}\Big]+\dots
\end{equation}
For real $\omega$ one has $\ln(-i\omega)=\ln|\omega|-i(\pi/2)\,{\rm sgn}\,\omega$, and the logarithm therefore carries an unavoidable imaginary part
\begin{equation}\label{eq:Imlog}
\omega\,{\rm Im}\,\Lambda(\omega)=-\frac{\pi}{2}\,\beta\,M^2\omega^2|\omega|\geq 0.
\end{equation}
This relation should be distinguished from the familiar one for black holes, where the running at ${\cal O}(\omega^2)$ appears to be tied to the dissipative coefficient at ${\cal O}(\omega)$. Indeed, for Schwarzschild we have  $d\lambda_2/d\ln\mu=-\eta r_s$ with tidal-heating coefficient $\eta=\tfrac{32}{45}M^6$~\cite{Rodriguez:2026iot}, whose absorptive part $\eta\omega$ is one power of $\omega$ lower than in Eq.~\eqref{eq:Imlog}. Equation~\eqref{eq:Imlog} is instead the analytic partner of the logarithm at the same order and, for a black hole, it is a subleading correction to tidal heating. Notice that the $\Lz$-independent part of $\beta$ coincides numerically with $-\eta r_s=-\tfrac{64}{45}M^7$, but it is present for horizonless stars as well~\cite{Apostolidis:2026qsg}, where the running is not sourced by horizon absorption. Indeed, the coefficient of the logarithm is fixed by the exterior geometry alone and does not depend on the boundary conditions imposed at the surface of the compact object. It is therefore universal, i.e.\ independent of the nature of the body, see for instance Ref. \cite{Garcia-Saenz:2026pcb}.

For a perfect-fluid star, whose absorption is negligible, Eq.~\eqref{eq:Imlog} is the leading absorptive term. Passivity is then consistent with, and indeed requires, $\beta\le0$, the ``asymptotically free'' sign found in Ref.~\cite{Apostolidis:2026qsg}. We stress that the absorptive part in Eq.~\eqref{eq:Imlog} is a tail effect of the full system rather than dissipation in the fluid, and that its presence relies on the far-zone completion~\eqref{eq:chilog}.

Under this assumption, Eq.~\eqref{eq:Imlog} corresponds, through Eq.~\eqref{eq:dispH}, to a spectral weight $d\rho(\omega')=|\beta|M^2\omega'\,d\omega'$ extending down to zero frequency for $\omega'\ll\omega_f$. A relativistic neutron star is then not strictly gapped, since besides its oscillation modes it carries a gravitational tail, whose contribution to the second moment in Eq.~\eqref{eq:moments} diverges logarithmically in the ultraviolet. This divergence is precisely the running of $\Lt$. At a scale $\mu<\omega_f$, $\Lt(\mu)$ counts the tail weight between $\mu$ and $\omega_f$ as if it lay above the gap, and the fixed-$\mu$ bound fails at $\mu r_s={\cal O}(10^{-2})$. Separating the weight below $\omega_f$ and matching the dispersion relation to Eq.~\eqref{eq:chilog} at small $\Omega$ identifies the modal contribution with the coupling evaluated at $\mu=\omega_f$, and leads to the refined bound (see Appendix~\ref{AppA} for a discussion)
\begin{equation}\label{eq:refined}
\Lt(\mu=\omega_f)\le\frac{\Lz}{(M\omega_f)^2}.
\end{equation}
This bound has a finite resolution, since the tail density is known only at leading order and the matching point is defined up to a factor of order unity. Shifting $\ln\mu$ by one unit changes $\omega_*/\omega_f$ by $\tfrac12|\beta|/\Lt\simeq2(M\omega_f)^2$, i.e.\ from $0.3\%$ at $\Lz=10^4$ to $5\%$ at $\Lz=5$.

\begin{figure}[t]
\includegraphics[width=\columnwidth]{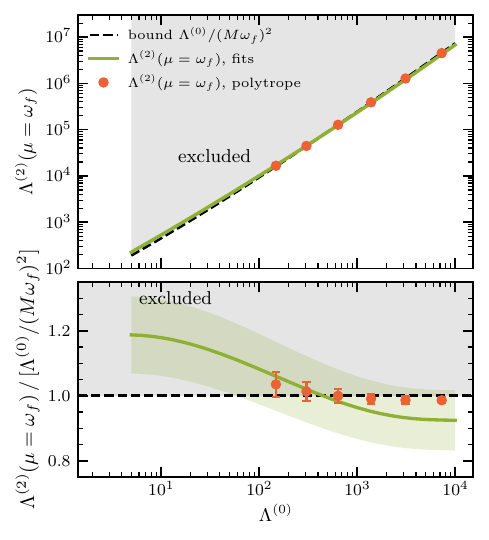}
\caption{Dynamical Love number evaluated at $\mu=\omega_f$ compared with the refined bound~\eqref{eq:refined} (dashed). Top: $\Lt(\mu=\omega_f)$ as a function of $\Lz$. Bottom: $\Lt(\mu=\omega_f)$ in units of the bound. The green line is the $\mu r_s=1$ fit of Ref.~\cite{Apostolidis:2026qsg} evolved to $\mu=\omega_f$ with Eq.~\eqref{eq:run}, with a band showing its $\pm10\%$ accuracy. Red points show the polytropic model of Ref.~\cite{Apostolidis:2026qsg} for $C=0.10$--$0.20$ (right to left), interpolated in $\ln\mu$ from their Fig.~5 to $\mu r_s=2M\omega_f$; error bars are the quadrature sum of the read-off uncertainty and of the change of $p$ under a unit shift of $\ln\mu$. Grey regions are excluded.
}
\label{fig:4}
\end{figure}

To quantify the comparison, we introduce the ratio
\begin{equation}\label{eq:p}
p\equiv\frac{\Lt(\mu=\omega_f)}{\Lz/(M\omega_f)^2}=\left[\frac{\omega_f}{\omega_*(\mu=\omega_f)}\right]^2 ,
\end{equation}
which is shown in the lower panel of Fig.~\ref{fig:4}, whereas the refined bound~\eqref{eq:refined} requires $p\le1$. For the polytropic model of Ref.~\cite{Apostolidis:2026qsg} we read off from their Fig.~5 the ratio $r\equiv\omega_*/\omega_f$ at the three plotted scales, $\mu r_s=0.01$, $0.1$ and $1$, and interpolate linearly in $\ln\mu$ to $\mu r_s=2M\omega_f$, so that $p=1/r^2$. Two sources of error affect $p$. The first is the read-off uncertainty $\delta r=\pm0.003$, which propagates to $\delta p_{\rm read}=2\,\delta r/r^3$. The second is intrinsic to the refined bound, since the matching point $\mu=\omega_f$ is defined only up to a factor of order unity, we take a unit shift of $\ln\mu$ as a measure of this ambiguity, which, by Eq.~\eqref{eq:run}, changes $p$ by $\delta p_{\rm res}=|\beta|(M\omega_f)^2/\Lz$. The error bars in Fig.~\ref{fig:4} are the quadrature sum
\begin{equation}\label{eq:dp}
\delta p=\left[\left(\frac{2\,\delta r}{r^3}\right)^2+\left(\frac{|\beta|(M\omega_f)^2}{\Lz}\right)^2\right]^{1/2},
\end{equation}
which ranges from $\delta p\simeq0.009$ at $C=0.10$ to $\delta p\simeq0.039$ at $C=0.20$ and is dominated by the second term for $C\gtrsim0.14$.

Evolving the $\mu r_s=1$ fit to $\mu=\omega_f$ with Eq.~\eqref{eq:run} gives $0.92\le\omega_*/\omega_f\le1.04$, with a deficit at the compact end comparable to the sum of the fit uncertainty and of the intrinsic resolution. A cleaner test is provided by the polytropic results of Ref.~\cite{Apostolidis:2026qsg}. Interpolating their Fig.~5 linearly in $\ln\mu$ between the curves that bracket $\mu r_s=2M\omega_f$, with $M\omega_f$ from the f--Love relation, we find that $\Lt(\mu=\omega_f)$ reaches $98.6\%$ of its maximum for $C\le0.12$, equals it at $C\simeq0.16$, and exceeds it by $1.3\%$ and $3.5\%$ at $C=0.18$ and $0.20$, see  Fig.~\ref{fig:4}, less than the intrinsic resolution of $\pm2.9\%$ and $\pm3.9\%$ at those compactnesses. The EFT-matched Love numbers are therefore consistent with the refined bound within its resolution, and nearly saturate it. As expected, the agreement improves for large $\Lz$, i.e.\ for small compactness.

Combined with an independent measurement of the f-mode frequency, the static deformability therefore provides an EoS-independent upper limit on the dynamical tide.

The normalizations of Sec.~\ref{sec:shy} can now be put on the same footing. Ref.~\cite{Apostolidis:2026qsg} gives the exact relation between their coupling and the PP one [their Eq.~(D1)], which in our variables reads
\begin{equation}
\Lt(\mu)=\Lt_{\rm PP}+25.86\,\Lz+0.90+\beta\ln(\mu r_s).
\end{equation}
The PP normalization is thus equivalent to evaluating the running coupling at $\ln(\mu r_s)\simeq6.3$, far above $\omega_f$, while the main-text normalization of Ref.~\cite{Saes:2025jvr} corresponds to $\ln(\mu r_s)\approx1$--$3$ at the level of the fits. The ordering of all the results in $\omega_*/\omega_f$, from the PP relation with the widest margin, through the main-text relation and Ref.~\cite{Apostolidis:2026qsg} at $\mu r_s=1$, down to Ref.~\cite{Apostolidis:2026qsg} at $\mu r_s=0.01$ below the fixed-scale bound, follows the ordering in $\mu$ dictated by $\beta<0$. 

\section{Conclusions}\label{sec:concl}
\noindent
In this paper we have confronted the Schwarz--Pick bound on the dynamical tidal response~\cite{Kehagias:2026opd} with the two most recent fully relativistic computations of the dynamical Love number of neutron stars. At the level of the two leading coefficients the bound is simple: $\Lt$ is positive and cannot exceed $\Lz/(M\omega_f)^2$ or, equivalently, $\omega_*\ge\omega_f$. The quasi-universal relations of Ref.~\cite{Saes:2025jvr} satisfy it for all the EoSs and compactnesses considered. With the normalization adopted in their main analysis, the relation lies within a few per cent of saturation, a quantitative confirmation of f-mode dominance in the relativistic regime, whereas with the PP normalization the bound is satisfied with a margin which grows with compactness. The difference between the two normalizations has exactly the form of a change of subtraction scale, and each of them corresponds to a fixed value of the renormalization scale in the EFT description.

For the running couplings of Ref.~\cite{Apostolidis:2026qsg} the bound is violated for sufficiently small values of $\mu r_s$. We have argued that this is to be expected. If the logarithmic running is completed in the far zone into $\ln(-i\omega)$, it comes with a gravitational tail whose spectral weight extends below $\omega_f$, and which the fixed-scale coupling counts as if it lay above the gap. The appropriate statement for a relativistic star is then the refined bound $\Lt(\mu=\omega_f)\le\Lz/(M\omega_f)^2$, which the available polytropic results satisfy within its intrinsic resolution of a few per cent. The same completion makes passivity consistent with the negative sign of the running found in Ref.~\cite{Apostolidis:2026qsg}, which differs from the black hole relation between running and tidal heating.

Our findings leave several directions worth pursuing. First, the comparison presented here relies on EoS-averaged fits, with scatter up to $5\%$ in $\Lt$ for the main-text relation and up to $\sim20\%$ for the PP one, on the f--Love relation, accurate to about $1\%$, and, for the refined bound, on values read off from published figures with an uncertainty of about $\pm0.003$. A sharper test requires the full frequency dependence of the response on the imaginary axis, which would allow a direct evaluation of the bracket~\eqref{eq:bracket}, or the values of $\Lt(\mu=\omega_f)$ computed star by star. 

Second, a derivation of the tail term from the full retarded response, beyond the near-zone matching, would put the refined bound on firmer ground. 

Third, in general relativity the f-mode is a quasinormal mode with quality factor $Q\sim10^3$, so that the spectral gap is sharp only up to corrections of order $1/Q$. 

Fourth, since the bound is saturated by a single mode, the combination of $\Lz$ with a determination of $\omega_f$, either independent or through the f--Love relation, provides an EoS-independent upper limit on $\Lt$ which may serve as a physical prior on the dynamical tide in parameter estimation. 

Finally, as already mentioned, it would be interesting to investigate whether the bound can be used to restrict the range of parameters describing the physics of neutron stars.
\begin{center}
{\bf Note Added}    
\end{center}
\noindent
After the completion of this work, we have become aware of a similar work \cite{Apostolidis:2026mvb}
Our results, when overlap is possible, agree with
theirs.
\begin{acknowledgments}
\noindent
It is a pleasure to  thank Valerio De Luca for comments on the draft and many interesting discussions. A.R.  acknowledges support from the Swiss National Science Foundation (project number CRSII5\_213497). The authors acknowledge the
use of  Anthropic’s Claude Code
as a supplementary research-assistance tool for implementing numerics, making figures, and editing prose. The scientific ideas and manuscript
writing are the work of the authors.
\end{acknowledgments}

\appendix

\section{Bound derivation}
\label{AppA}
\noindent
Let us consider the retarded dispersion representation of the retarded response $\Lambda(\omega)$ for $\operatorname{Im}\omega\geq 0$ 
\begin{equation}
    \label{A1}
\Lambda(\omega)=\int_0^\infty \frac{\omega'^2}{\omega'^2-\omega^2}d \rho(\omega'),
\end{equation}
with a finite static moment $\Lambda^{(0)}=\int_0^\infty d\rho(\omega')$ and no other polynomial substraction terms. We assume that the measure $\rho$ has a smooth density behaviour in  $0\leq \omega\leq \omega_f$ so that 
\begin{align}
    \frac{d\rho}{d\omega}&=bM^2\omega+r(\omega), \qquad b>0,\nonumber \\
    r(\omega)&=\mathcal{O}(\omega^{1+\delta}), \qquad\qquad\hskip 0.05cm \delta>1, \label{eq:rho'}
\end{align}
as $\omega\to 0$, and in addition, $r(\omega)/\omega^2$ is integrable, whereas the coefficient $b$ of the linear term of $\rho'$ in (\ref{eq:rho'}) must be independently matched to the EFT beta function through $b=-\beta$. Now, on the imaginary axis with $\omega=i \Omega$
 we get from Eq. \eqref{eq:rho'} 
 \begin{align}
     \frac{\Lambda^{(0)}-\Lambda(i\Omega)}{M^2 \Omega^2}&=\frac{1}{M^2}\int_0^\infty 
     \frac{d\rho(\omega')}{\omega'^2+\Omega^2}\nonumber\\ 
     &=b\int_0^{\omega_f} \frac{\omega'd\omega'}{\omega'^2+\Omega^2}+\frac{1}{M^2}\int_0^{\omega_f} 
     \frac{r(\omega')d\omega'}{\omega'^2+\Omega^2}\nonumber \\
     &+\frac{1}{M^2}\int_{\omega_f}^\infty
     \frac{d\rho(\omega')}{\omega'^2+\Omega^2}.
 \end{align}
 In the low frequency limit $|\Omega|\ll \omega_f$, we find 
 \begin{align}
\frac{1}{M^2}\int_0^\infty
\frac{d\rho(\omega')}{\omega'^2+\Omega^2}
&=\frac{b}{2}\ln
\frac{\omega_f^2+\Omega^2}{\Omega^2}
+I_f+C_f+o(1)\nonumber\\
&=b\ln\frac{\omega_f}{\Omega}
+I_f+C_f,
\end{align}
 where 
 \begin{align}
     I_f&=\frac{1}{M^2}\int_{\omega_f}^\infty \frac{d\rho(\omega')}{\omega'^2}, \nonumber \\
C_f&=\frac{1}{M^2}\int_0^{\omega_f}\frac{r(\omega')d\omega'}{\omega'^2},
 \end{align}
 and therefore
 \begin{equation}
     \frac{\Lambda^{(0)}-\Lambda(i\Omega)}{M^2 \Omega^2}=b\ln\frac{\omega_f}{\Omega}+I_f+C_f, \qquad \Omega\ll \omega_f.
 \end{equation}
 Hence, the spectral coefficient  at a substraction scale $\mu$ is 
 \begin{equation}
 \Lambda^{(2)}=I_f+C_f+b\ln \frac{\omega_f}{\mu}, \label{L2}
 \end{equation}
 so that we get Eq. \eqref{eq:chilog}, i.e. 
 \begin{equation}\label{eq:chilogap}
\Lambda(\omega)=\Lz+(M\omega)^2\Big[\Lt(\mu)+\beta\ln\frac{-i(\omega+i0)}{\mu}\Big]+\dots,
\end{equation}
 by using $\Omega=-i\omega$ and the identification $b=-\beta$. 
At $\mu=\omega_f$ the explicit logarithm in Eq. \eqref{L2} vanish 
so that we get 
\begin{equation}
    \Lambda^{(2)}(\mu=\omega_f)=I_f+C_f, \label{L21}
\end{equation}
Since for the part of the positive measure above $\omega_f$ we have the inequality 
\begin{equation}
    0\leq \int_{\omega_f}^\infty \frac{d\rho(\omega')}{\omega'^2}\leq\frac{1}{\omega_f^2}\int_{\omega_f}^\infty d\rho(\omega')\leq\frac{\Lambda^{(0)}}{\omega_f^2},
\end{equation}
we get that 
\begin{equation}
    0\leq I_f\leq \frac{\Lambda^{(0)}}{M^2\omega_f^2}, 
\end{equation}
and therefore
\begin{equation}
    \Lambda^{(2)}(\mu=\omega_f)-C_f\leq 
    \frac{\Lambda^{(0)}}{M^2\omega_f^2}. 
\end{equation}
The largest frequency one may consider in the binary merger of two neutron stars is the    contact frequency, that is the inspiral frequency of the GW when the 
neutron stars come into contact, $\omega_{c}\sim 0.7(C/0.1)^{3/2}$ kHz, where we have taken the typical mass of the  neutron stars in the binary to be 1.4 $M_\odot$. For the same mass,  $\omega_f\simeq 0.03 (C/0.1)^{3/2}$ kHz \cite{Andersson:1997eq} and  the low-frequency expansion is trustable.

A  stronger result follows at leading  Newtonian order with dissipation neglected. Let us consider a stable, nonrotating barotropic perfect fluid star with smooth equation of state where its perturbations obey the same equation of state as the equilibrium configuration. Then, the lowest frequency among the modes with a non-zero quadrupolar tidal overlap is the fundamental frequency $\omega_f$, whereas the remaining coupled modes are pressure overtones with $\omega>\omega_f$.
 If we neglect damping, the Newtonian perturbations admit an expansion in normal modes, and the corresponding spectral measure is 
 \begin{equation}
     d \rho(\omega)=\sum_n a_n \delta(\omega-\omega_n) d\omega \, \qquad a_n\geq 0,
 \end{equation}
where the sum runs over all quadropolar normal modes. Their weights $a_n$ are nonnegative because they are proportional to squared tidal overlaps divided by positive mode inertias and squared mode frequencies~\cite{Pitre:2023xsr}. 
 Then, using Eq. \eqref{A1}, we find that 
 \begin{equation}
     \Lambda(\omega)=\sum_n \frac{a_n \omega_n^2}{\omega_n^2-\omega^2},
 \end{equation}
 which after expanding it for $|\omega|\ll \omega_f$, gives
\begin{equation}
    \Lambda(\omega)=\Lambda^{(0)}+(M\omega)^2 \Lambda^{(2)}+\mathcal{O}(\omega^4),
\end{equation}
where 
\begin{equation}
    \Lambda^{(0)}=\sum_n a_n, \qquad \Lambda^{(2)}=\frac{1}{M^2}\sum_n \frac{a_n}{\omega_n^2}.
\end{equation}
Then, since $a_n\geq 0$ and $\omega_n\geq \omega_f$, we get 
\begin{equation}
    0\leq \Lambda^{(2)}\leq \frac{1}{M^2\omega_f^2}\sum_na_n=\frac{\Lambda^{(0)}}{(M\omega_f)^2},
\end{equation}
which establishes Eq. \eqref{eq:refined} at leading Newtonian order with no dissipation. The f-mode typically dominates the tidal response because it couples strongly to the quadrupole tidal field, whereas the pressure overtones have smaller overlaps. The bound is nearly saturated when the f-mode accounts for almost all the static response.

Our numerical results from Fig. \ref{fig:4} indicate that the relativistic corrections encoded in the coefficient $C_f$ are tiny in the limit of small compactness.

\bibliography{refs}

\end{document}